\documentclass[preprint,amsmath,amssymb,aps]{revtex4-1}
\usepackage{graphicx}
\usepackage{bm}
\usepackage{hyperref}
\usepackage[mathlines]{lineno}
\usepackage{geometry}

\begin{document}

\title{Chiral Triclinic Metamaterial Crystals Supporting Isotropic Acoustical Activity and Isotropic Chiral Phonons}

\author{Yi Chen$^{1}$}
\email{yi.chen@partner.kit.edu; chenyi221@gmail.com}
\author{Muamer Kadic$^{2}$}
\author{Martin Wegener$^{1,3}$}
\affiliation{$^{1}$Institute of Applied Physics, Karlsruhe Institute of Technology (KIT), 76128 Karlsruhe, Germany}
\affiliation{$^{2}$Institut FEMTO-ST, UMR 6174, CNRS, Universit\'{e} de Bourgogne Franche-Comt\'{e}, 25000 Besan\c{c}on, France}
\affiliation{$^{3}$Institute of Nanotechnology, Karlsruhe Institute of Technology (KIT), 76021 Karlsruhe, Germany}

\date{\today}

\begin{abstract}
Recent work predicted the existence of isotropic chiral phonon dispersion relations of the lowest bands connected to isotropic acoustical activity in cubic crystalline approximants of 3D chiral icosahedral metamaterial quasicrystals. While these architectures are fairly broadband and presumably robust against fabrication tolerances due to orientation averaging, they are extremely complex, very hard to manufacture experimentally, and they show effects which are about an order of magnitude smaller compared to those of ordinary highly anisotropic chiral cubic metamaterial crystals. Here, we propose and analyze a chiral triclinic metamaterial crystal exhibiting broadband isotropic acoustical activity. These 3D truss lattices are much less complex and exhibit substantially larger effects than the 3D quasicrystals at the price of being somewhat more susceptible to fabrication tolerances. This susceptibility originates from the fact that we have tailored the lowest two transverse phonon bands to exhibit an “accidental” degeneracy in momentum space.
\end{abstract}

\keywords{chirality; metamaterials; isotropic acoustical activity; chiral phonons; accidental degeneracy}


\maketitle

\section{Introduction}
The properties of elastic waves in ordinary crystals \cite{Authier2003} and in rationally designed artificial crystals called metamaterials \cite{milton2002, ranganathan2008universal, fang2019energy} are generally highly anisotropic, even in the long-wavelength limit. However, isotropic behavior in (artificial) crystals is not forbidden by any law of physics. It is instructive to consider the analogy to quantum mechanics, where energy levels can be degenerate by symmetry or due to some other fundamental principle. If no such principle applies, energy levels can still coincide “accidentally” or by designing them to be degenerate. 

Isotropic elastic behavior has been studied for achiral artificial architectures \cite{messner2016optimal, berger2017mechanical} and more recently also for chiral metamaterials \cite{duan2018predictive, Kadic2019b, chen2020isotropic}. Chiral crystals allow behaviors that are generally forbidden in the absence of chirality. For example, chirality in three dimensions enables strain-to-twist conversion in the static regime \cite{Frenzel2017, chen2020high} and chiral phonons \cite{zhu2018observation} leading to acoustical activity \cite{portigal1968acoustical, Frenzel2019} in the dynamic regime. Three-dimensional (3D) chiral phonons mean that the average displacement vector of one crystal unit cell orbits in circles around its rest position, clockwise or counter-clockwise. The orbit plane is perpendicular to the phonon wave vector. Along the direction of the phonon wave vector, these orbits are phase delayed, such that a snapshot of the chiral phonon displacements forms a helix. This helix propagates along the direction of the phonon wave vector with the associated acoustical phase velocity. Therefore, under these conditions, linearly polarized transverse acoustical (TA) phonons are no longer eigenstates of the system. They can be decomposed into left- and right-handed chiral-phonon eigenstates, which propagate with different phase velocities, leading to a rotation of an incident linear polarization versus propagation distance. This phenomenon has been called acoustical activity \cite{portigal1968acoustical, pine1970direct}, in close analogy to optical activity \cite{thomson2010baltimore}. It can be used and applied to manipulate the polarization of elastic waves.

A sufficient symmetry requirement for the occurrence of chiral phonons in 3D is chirality of the crystal lattice combined with the axis of the phonon wave vector having three-fold or higher rotational symmetry \cite{fernandez2013forward}. However, as we will show by constructive examples in this paper, this sufficient symmetry requirement is not a necessary requirement. Likewise, in planar geometries, 2D chiral phonons have been shown to occur in achiral honeycomb crystal lattices for special in-plane phonon wave vectors \cite{zhu2018observation, chen2019topologicalphase}. 

Our previous work on approximants of 3D quasi-crystalline chiral mechanical metamaterials has already shown that 3D cubic crystal unit cells can be constructed that effectively and approximately lead to isotropic acoustical activity \cite{chen2020isotropic}. However, these truss lattices have been extremely complex with more than one hundred thousand ordinary elastic rods in a single cubic approximant unit cell. Hence, these architectures are extremely hard to manufacture, and presently even out of reach experimentally. 

Therefore, in this paper, we rationally design largely simplified artificial cubic crystals leading to isotropic chiral phonons and isotropic acoustical activity over a broad frequency range. In addition to being much simpler, these architectures show a much larger relative frequency splitting of the two lowest acoustical bands, associated to larger polarization rotation effects due to acoustical activity. As we exploit “accidental” degeneracies for the 3D truss-lattice architectures discussed here, the price we have to pay is that they are not as robust against deviations from the ideal design parameters as our previous 3D quasi-crystal approximants, which led to isotropic behavior due to full rotational symmetry on average. Nevertheless, we show that the 3D crystals presented here are reasonably robust. Overall, our novel design blueprints bring us much closer to attractive future experimental realizations by advanced 3D additive manufacturing on the micrometer scale.

\section{Rational design process}

Many material aspects in crystals, such as thermal, acoustical or optical properties, are described by second-order tensors \cite{Authier2003}. In that case, spatial symmetry alone, such as cubic symmetry, guarantees isotrop properties. In sharp contrast, elastic properties are encapsulated in a fourth-order tensor. As a result, isotropy cannot be obtained simply from crystal symmetry, neither in the chiral nor in the achiral case \cite{Authier2003}. In this paper, we construct an isotropic chiral metamaterial crystal with broadband isotropic acoustical activity through accidental degeneracy. Our rational design process can be summarized by three main steps: First, we identify a simple achiral truss lattice with ordinary rods (cf. Fig.\,1(a)) which supports achiral isotropic elasticity. Second, a chiral version of the truss lattice (cf. Fig.\,1(c)) is obtained by replacing all of the rods (cf. Fig.\,1(a)) with designed chiral meta-rods (cf. Fig.\,1(b)). Third, we achieve accidental isotropic acoustical activity through optimizing the geometry of the three different groups of meta-rods. The details are described in what follows.

\begin{figure}[h!]
\begin{center}
\includegraphics[width=15cm, angle=0]{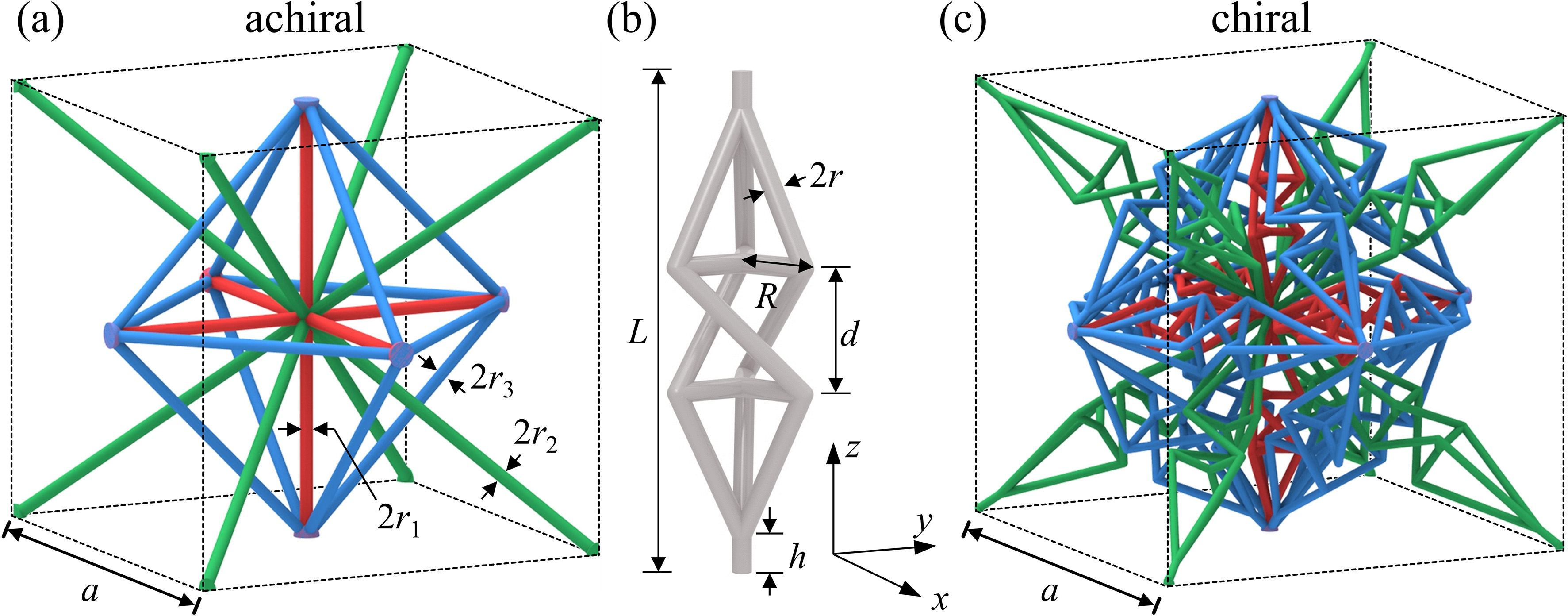}
\caption{(a) Illustration of a three-dimensional (3D) cubic achiral unit cell assembled from ordinary rods with circular cross section. Each unit cell contains 6 rods (colored in red), 8 rods (colored in green) and 12 rods (colored in blue) along the cubic principal directions, the space diagonal directions, and the face diagonal directions, respectively. A metamaterial crystal is obtained by placing the unit cell on a simple cubic lattice with lattice constant $a$. In this paper, to be specific, we choose $a = 200\,\mathrm{\mu m}$ throughout. However, the resulting frequencies can easily be scaled to other choices. We achieve accidental elastic isotropy with the relative diameters $2r_1/a = 0.04, 2r_2/a = 0.026$, and $2r_3/a = 0.03$, for the red, green and blue rods, respectively. (b) Designed chiral meta-rod assembled from rods of the same diameter $2r$. Other geometry parameters are the length $L$, the radius $R$, the distance $d$, and the length of the end part $h$. The chiral meta-rod exhibits three-fold rotational symmetry along its axial direction. (c) A 3D cubic chiral unit cell is obtained by replacing each of the red, green, and blue ordinary rods in (a) by the chiral meta-rod. The overall lengths $L$ of the meta-chiral rods along the cubic principal directions, the space diagonal directions, and the face diagonal directions are scaled accordingly. Again, we achieve accidental isotropic chiral elasticity with the optimized parameters $d/L= 0.248$, $R/L = 0.143$, $d/L = 0.150$, $R/L = 0.112$, and $d/L = 0.150$, $R/L = 0.143$, for meta-rods along the cubic principal directions, the space diagonal directions and the face diagonal directions, respectively. The other parameters are $2r/a = 0.02$ and $2h/(L-d) = 0.02$. For all of the rods, we choose the same constituent material parameters: Young’s modulus $E=4.18\,\mathrm{GPa}$, Poisson’s ratio $\nu=0.4$, and mass density $\rho =1.15\,\mathrm{g/}\mathrm{cm}^3$.}
\label{Figure1}
\end{center}
\end{figure}

We start by selecting the simple-cubic lattice with lattice constant $a$ as the translational lattice for our metamaterial crystal. This choice is motivated by its simplicity. To arrive at a stable truss structure, we introduce 6 rods of the same diameter $2r_1$ along the principal cubic directions (cf. red rods in Fig.\,1(a)). A cubic crystal is obtained by tessellation of the cubic unit cell along the cubic principal directions. The cubic crystal has four axes with three-fold rotational symmetry  and four axes with three-fold rotational symmetry. By symmetry \cite{Authier2003}, the cubic crystal needs three independent effective elastic constants, $C_{11}$, $C_{12}$, and $C_{44}$ (in Voigt notation) \cite{lai2009introduction}, whereas an isotropic elastic material only requires two independent elastic constants with the “isotropy” constraint $C_{44} = (C_{11} – C_{12})/2$. In order to achieve isotropic elastic behavior, we further connect the body center and the eight cubic corners of the unit cell by rods with diameter $2r_2$ (cf. green rods in Fig.\,1(a)). By optimizing the two diameters, $2r_1$ and $2r_2$, the three effective elastic constants, $C_{11}$, $C_{12}$ and $C_{44}$, for the cubic crystal can be tailored to accidentally satisfy the isotropy constraint. In our numerical optimization, this constraint is imposed by demanding degeneracy between the first and the second transverse bands along a face diagonal direction of the cubic crystal in the long-wavelength limit. To make the desired isotropic elastic behavior more robust, we increase the coordination number at the face centers by introducing a third type of rods, with diameter $2r_3$, connecting the centers of adjacent faces (cf. blue rods in Fig.\,1(a)). As a compromise between large coordination number and simplicity of the lattice, we do not connect the six face centers and the eight cubic corners. As an example, we obtain an isotropic achiral crystal with the optimized relative diameters, $2r_1/a = 0.04$, $2r_2/a = 0.026$, and $r_3/a = 0.03$. 

To arrive at a chiral unit cell (cf. Fig.\,1(c)), we simply replace the above three types of ordinary rods (cf. red, green, and blue rods Fig.\,1(a)) by the previously mentioned chiral meta-rods (Fig.\,1(b)). The same type of truss-based chiral meta-rods has been used in our previous work \cite{chen2020isotropic}. A chiral metamaterial crystal is thus obtained similarly by periodically replicating the cubic chiral unit cell in space. In contrast to achieving isotropy through orientational averaging in our previous design based on cubic approximants of a quasi-crystalline lattice \cite{chen2020isotropic}, the resulting chiral metamaterial crystal behavior cannot automatically be isotropic. We need to optimize the chiral meta-rods to obtain isotropic acoustical activity through accidental degeneracy. Six dimensionless variables, namely the two ratios $d/L$ and $R/L$ for the three types of chiral meta-rods, are considered in the optimization. These two ratios are the most important factors for the chiral effect of the meta-rods. The other parameters of the meta-rods are fixed. We use a normalized rod diameter of $2r/a = 0.02$ (leading to the same diameter of all rods with respect to the unit cell size $a$) and $2h/(L-d) = 0.2$ (for practical reasons)\cite{chen2020isotropic}. In the numerical optimization, we aim at the same nonzero relative frequency splitting between the first and the second transverse band for elastic waves propagating along three specific directions, i.e, a selected principal cubic direction (001), a selected face diagonal direction (110), and a selected space diagonal direction (111). Technically, this six-dimensional optimization is performed by using the function “fminsearch” from the commercial Matlab program package. We achieve accidental isotropic acoustical activity with the following optimized geometric parameters: $d/L= 0.248$, $R/L = 0.143$, $d/L = 0.150$, $R/L = 0.113$, and $d/L = 0.150$, $R/L = 0.143$, for the red, green, and blue meta-rods, respectively. Let us note that the chiral unit cell is comprised of a total of only 26 chiral meta-rods. This number is around 3 orders of magnitude smaller compared to the 16768 chiral meta-rods in our previous design based on chiral quasicrystal approximants \cite{chen2020isotropic}. 

We emphasize that the symmetry of the resulting chiral crystal depicted in Fig.\,1(c) is low. The blue part and the green part (cf. Fig.\,1(c)) both show four-fold rotational symmetry along the three cubic principal directions and three-fold rotational symmetry along the four space diagonal directions, whereas the red part does not have any nontrivial rotational symmetry. Strictly speaking, the designed metamaterial structure is a triclinic crystal without any rotation or mirror symmetries except for the simple-cubic translational symmetry \cite{hahn1983international, mehl2017aflow}. However, our following calculations indicate that the properties of the metamaterial crystal are very nearly isotropic, i.e., they are even more symmetric than those of a general cubic crystal. In addition, we have time-inversion symmetry, which means that when replacing ${\bf k}\rightarrow -{\bf k}$ and flipping the handedness of the mode, the eigenfrequency stays the same. We explout this symmetry in our calculations to reduce the numerical effort. In particular, the calculated eigenfrequencies and eigenmodes show four-fold rotational symmetry along all cubic principal directions and three-fold rotational symmetry along all space diagonal directions. This finding (see below) will {\it a posteriori} justify our ansatz in which we only consider a single cubic direction, a single face diagonal direction, and a single space diagonal direction for optimizing the relative frequency splitting.

We solve the elastodynamic equation $-\rho \omega^2 {\bf u} = E/(2+2\nu)/(1-2\nu) \nabla (\nabla \cdot {\bf u}) +E/(2+2\nu) \nabla ^2 {\bf u}$ for the achiral and chiral metamaterial to obtain the phonon band structure and corresponding eigenmodes. $\omega = 2\pi f$ and ${\bf u}$ represent the angular eigenfrequency and the displacement field, respectively. Simple-cubic Bloch periodic boundary conditions are applied to the six surfaces of the unit cell. All other boundaries of the rods are traction free. To reduce the numerical effort, we explicitly exploit time-inversion symmetry in our calculations. Time-inversion symmetry means that when replacing ${\bf k}\rightarrow -{\bf k}$ and simultaneously flipping the handedness of the mode, the eigenfrequency stays the same. 
The elasticity equation is solved using a finite-element approach implemented within the commercial software COMSOL Multiphysics\textsuperscript{\textregistered} (MUMPS solver) \cite{zienkiewicz2000finite}. To obtain a smooth geometry at the junctions of all cylindrical rods, i.e., at the body center, the six face centers, and the eight corners of the cubic unit cell, we add a small sphere at each of these points. Its diameter is $4r_1$ for the achiral case and $4r$ for the chiral case, respectively. The chiral (achiral) unit cell is discretized into around 270 (33) million tetrahedral meshes, with the maximum size being one third of the minimal rod radius. For all rods in all of the metamaterial crystals, we choose the Young’s modulus $E=4.18\,\mathrm{GPa}$, the Poisson’s ratio $\nu=0.4$, and the mass density $\rho =1.15\,\mathrm{g\,}{\mathrm{cm}}^{\mathrm{-}\mathrm{3}}$. These parameters correspond to a typical polymer as, for example, used in \cite{Frenzel2019} at frequencies around $100\,\rm kHz$. Results for other Young's moduli and mass densities at fixed Poisson's ratio can easily be obtained by scaling the results of the following section. For example, increasing only the Young's modulus by a certain factor will increase all eigenfrequencies by the square root of this factor.

\section{Numerical results and discussion}

We start by showing results for the optimized achiral cubic crystal. Figure 2(a)-(c) exhibit the calculated phonon dispersion relations for elastic waves propagating along a cubic principal direction, a face diagonal direction, and a space diagonal direction, respectively. As expected for an isotropic elastic material, the two transverse bands are degenerate along these three directions. For the high-symmetry cubic principal direction and the space diagonal direction, the two transverse bands overlap over the entire wave vector range by symmetry, while the degeneracy along the face diagonal direction is accidental in the sense discused above (cf. Fig.\,2(b)). Precisely, this degeneracy has been enforced by our optimized parameter choice. The relative frequency splitting between the two transverse bands, $2(f_2-f_1)/(f_2+f_1)$, along the face diagonal direction is not larger than $0.04\%$ for any wave number. Furthermore, inspection of the eigenmodes reveals that the first two transverse eigenmodes are linearly polarized (not depicted), as to be expected.

\begin{figure}[htbp]
\begin{center}
\includegraphics[width=15.00cm, angle=0]{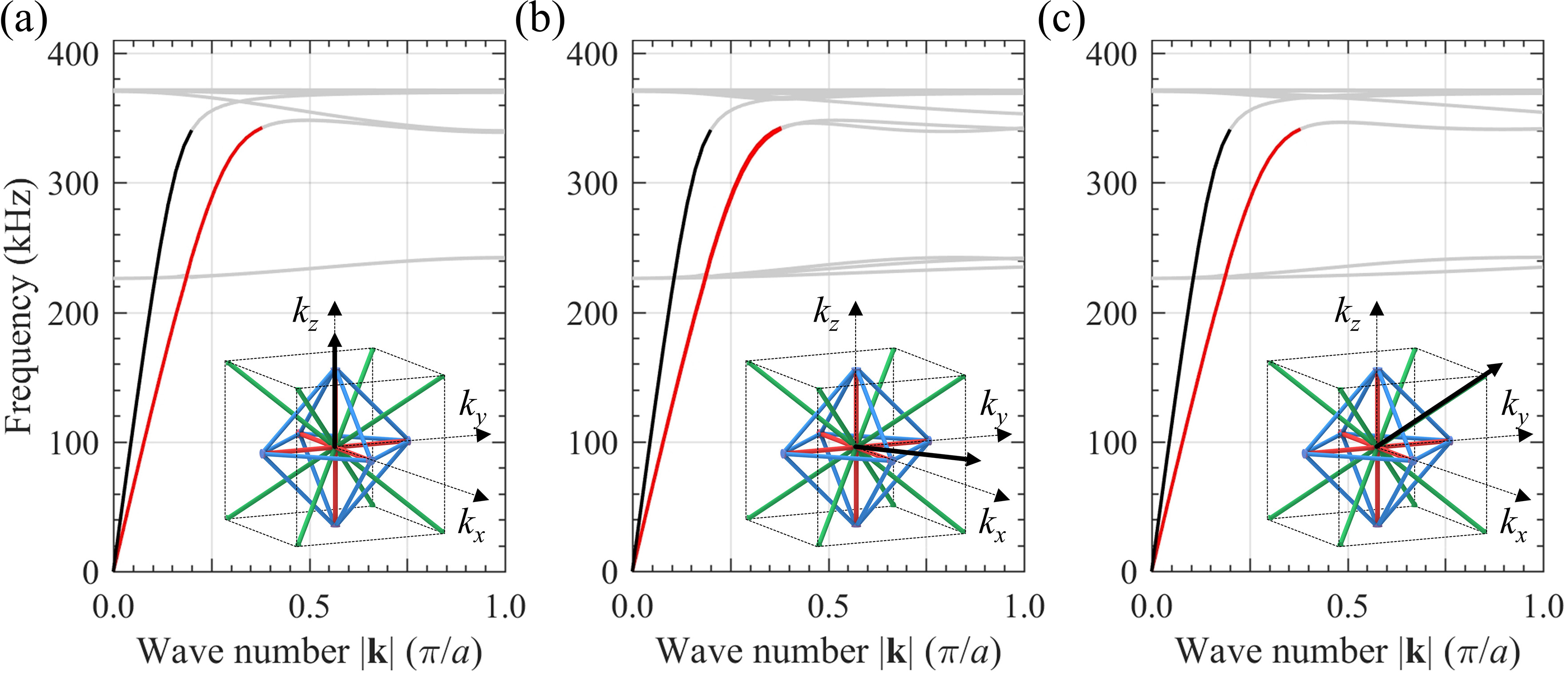}
\caption{Numerically calculated phonon dispersion relations for elastic waves propagating in the cubic achiral metamaterial crystal along: (a) a cubic principal direction (001), (b) a face diagonal direction (110), and (c) a space diagonal direction (111). The thick black arrow in each inset denotes the corresponding wave vector direction. The transverse and longitudinal bands are colored in red and black, respectively. Higher bands, which are less interesting for the purpose of this paper, are plotted in gray. As expected for an effective medium with isotropic properties, the two transverse bands (red) are degenerate in the long-wavelength or low-frequency limit and cannot be distinguished within the line thickness.}
\label{Figure2}
\end{center}
\end{figure}

\begin{figure}[htbp]
\begin{center}
\includegraphics[width=15.00cm, angle=0]{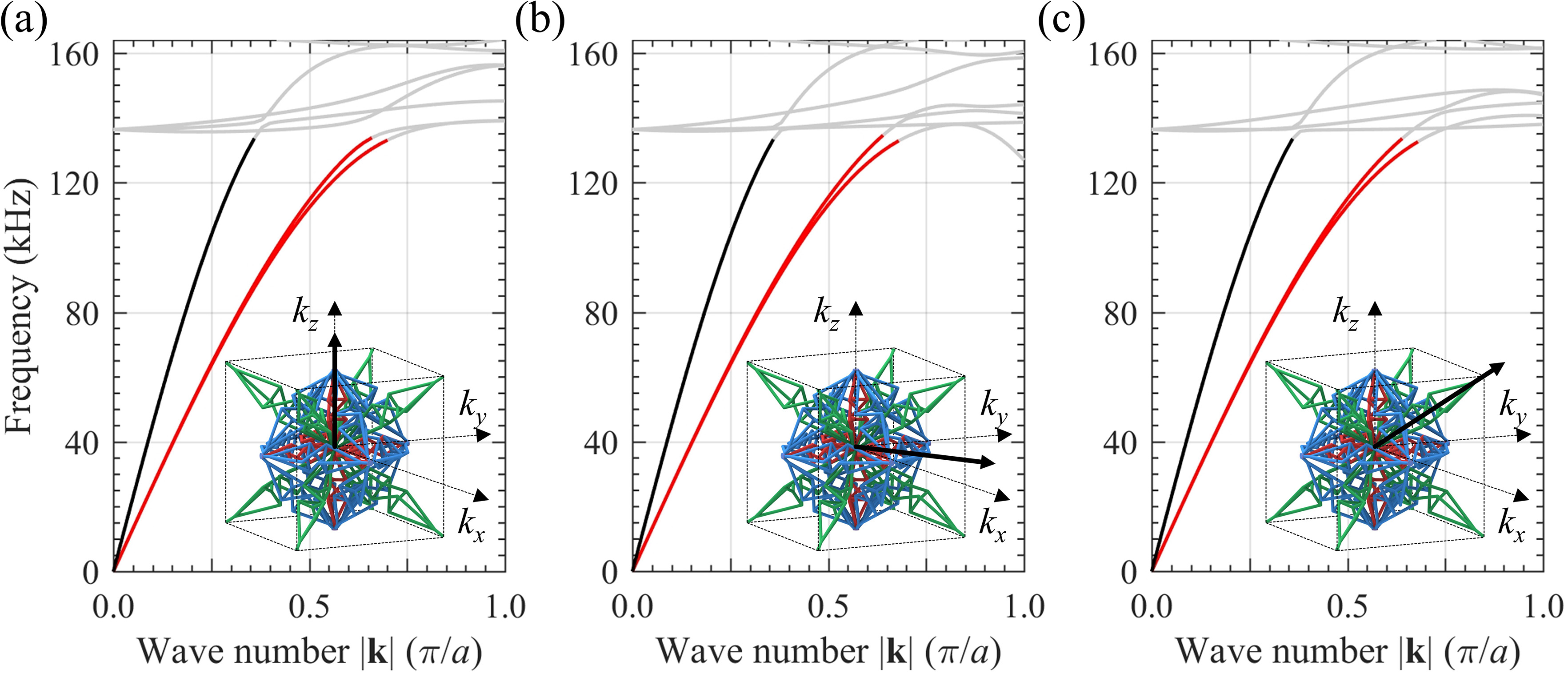}
\caption{Same as Fig.\,2, but for the chiral metamaterial crystal. (a) a cubic principal direction (001), (b) a face diagonal direction (110), and (c) a space diagonal direction (111). Note that the transverse (red) dispersion relations and their splitting are nearly direction independent, indicating an isotropic behavior, including isotropic acoustical activity.}
\label{Figure3}
\end{center}
\end{figure}

Next, we present in Fig.\,3(a)-(c) the calculated phonon bands for the designed chiral metamaterial crystal. Here, a significant splitting between the lowest two transverse bands (red) occurs, which becomes especially visible at large frequencies. However, the individual transverse bands are very nearly identical for all three depicted propagation directions. This chirality induced frequency splitting leads to circularly polarized transverse eigenmodes. To illustrate this important aspect, we plot in Fig.\,4(a)-(c) the two transverse eigenmodes for a fixed length of the wave vector, i.e., for $|\mathbf{k}| = 1/2(\pi/a)$, along the three aforementioned directions. The first (second) transverse eigenmodes are left (right) circularly polarized, i.e., the rotation direction of the unit cell center of mass and the wave vector follows a left (right) hand rule. Refer to Supplemental Videos for animations of the circularly polarized eigenmodes. Corresponding to the different slopes of the two transverse dispersion branches, the left and right circularly polarized eigenmodes propagate at different phase velocities. As discussed in the introduction, this difference leads to the phenomenon of acoustical activity. Therefore, the polarization axis of a linearly polarized transverse wave will rotate by a angle $\Delta k\,\l/2$ after propagating over a distance $\l$, where $\Delta k$ denotes the difference of the wave numbers corresponding to the left and right circularly transverse eigenmodes at a given frequency. For the designed chiral metamaterial crystal, the rotation angle is about $7.3^\circ$ per unit cell at a frequency about $130\,\rm{kHz}$, which is comparable in magnitude to the rotation power of previously designed chiral cubic metamaterials with highly anisotropic properties \cite{Frenzel2019}. 

\begin{figure}[htbp]
\begin{center}
\includegraphics[width=14.0cm, angle=0]{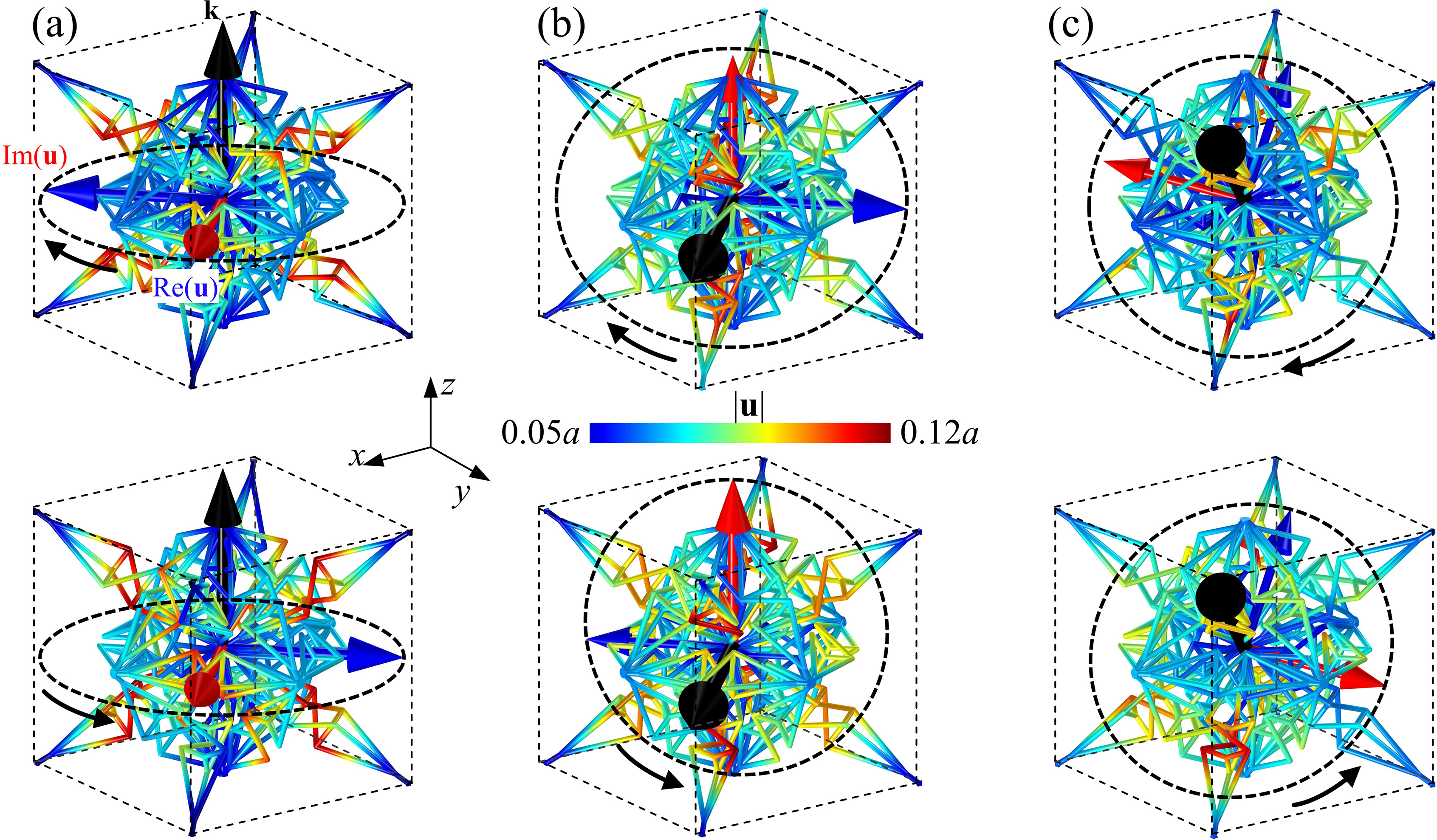}
\caption{Transverse eigenmodes corresponding to the three wave vectors: (a) $\mathbf{k} = \pi/(2a) (0, 0, 1)$, (b) $\pi/(2a) (0, 0, 1)$, and (c) $\pi/(2a) (0, 0, 1)$. The first (second) row corresponds to the first (second) transverse eigenmode. The false-color scale represents the absolute value of the displacement, normalized to the lattice constant $a$. For clarity, the magnitude of the displacements has been largely exaggerated. The red (blue) arrow is proportional to the real (imaginary) part of the displacement vector averaged over one unit cell. The phonon wave vector is indicated by the black arrow. The real and imaginary part of the displacement vector have the same length and they are perpendicular to each other in a plane orthogonal to the wave vector. This behavior is equivalent to circularly polarized transverse eigenmodes. This means that, for both eigenmodes, the unit cell rotates along a circular trajectory (see dashed black circles and small arrows) around its rest position in a plane perpendicular to the wave vector. The first (second) eigenmode is left (right) circularly polarized with respect to the wave vector direction. Refer to Supplemental Videos for animations.}
\label{Figure4}
\end{center}
\end{figure}

\begin{figure}[htbp]
\begin{center}
\includegraphics[width=15.0cm, angle=0]{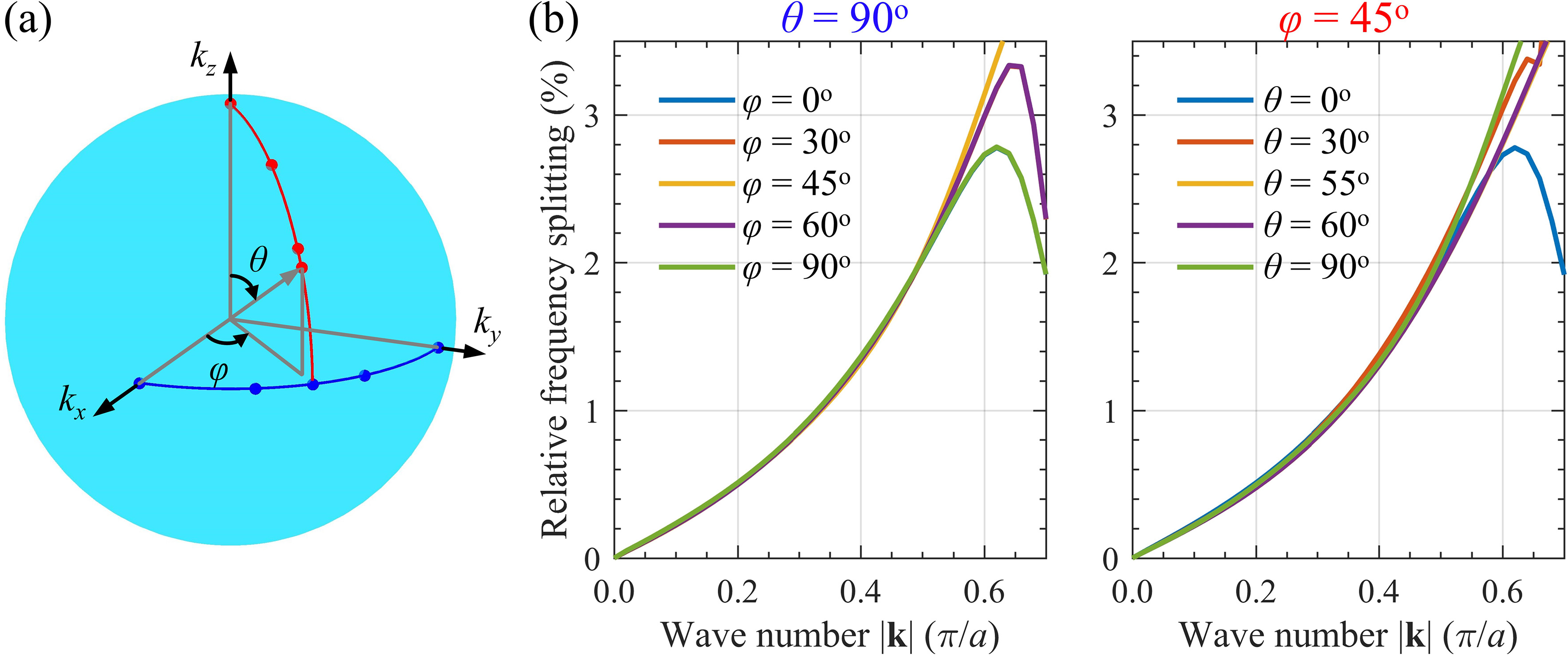}
\caption{(a) Representation of the polar angle $\theta$ and the azimuthal angle $\varphi$ for denoting the wave propagating directions. (b) Relative frequency splitting between the two transverse bands, $2(f_2-f_1)/(f_2+f_1)$ versus wave number for fixed $\theta=90^{\rm o}$ (left) and fixed $\varphi=45^{\rm o}$ (right), respectively. The relative frequency splitting is nearly independent of $\varphi$ and $\theta$ and vanishes in the long-wavelength limit, consistent with isotropic acoustical activity.}
\label{Figure5}
\end{center}
\end{figure}

\begin{figure}[htbp]
\begin{center}
\includegraphics[width=15.0cm, angle=0]{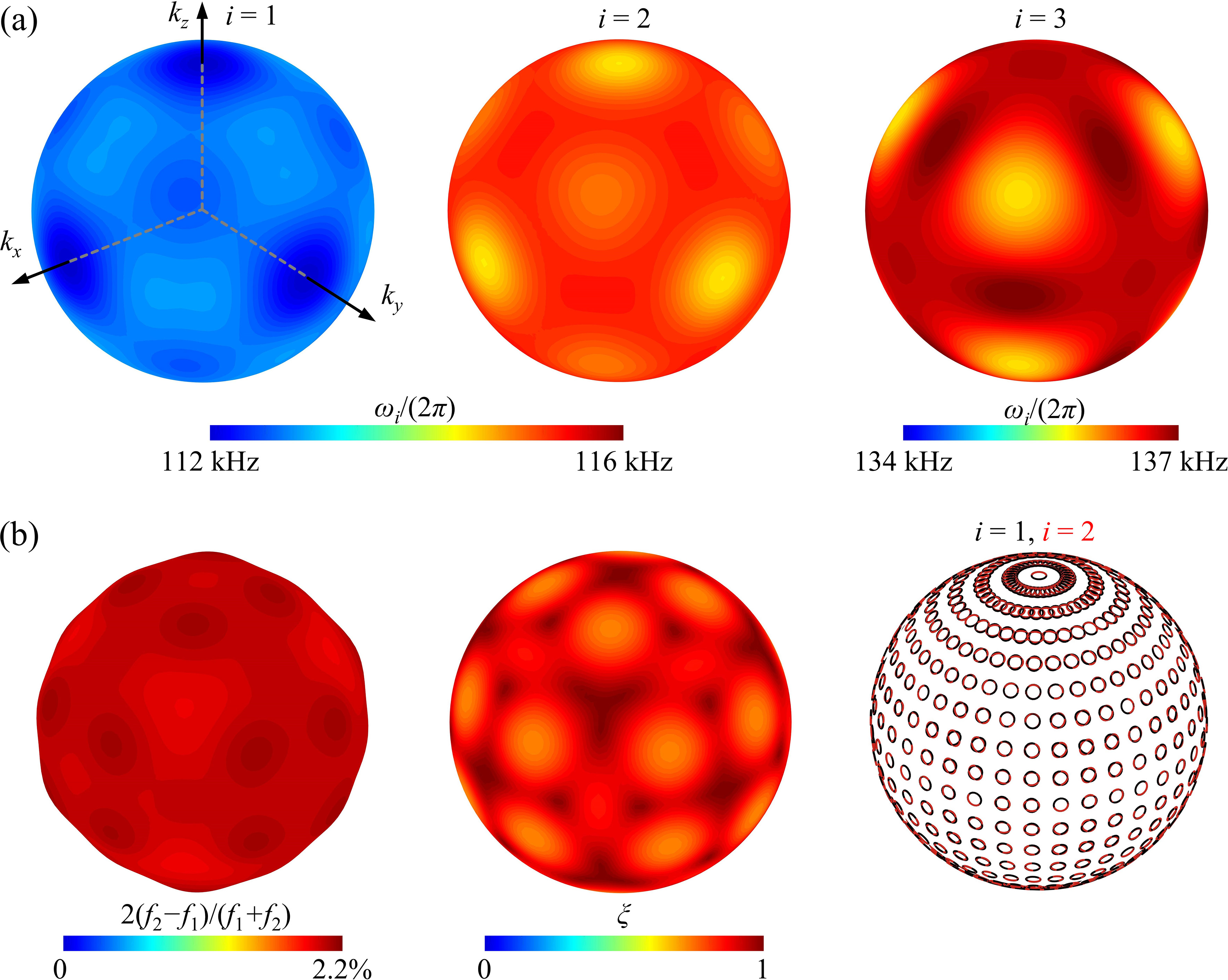}
\caption{(a) Direction dependence of the eigenfrequencies $f_i = \omega_i / (2\pi)$ for the three lowest acoustic bands with $i = 1,\,2,\,3$ for a fixed length of the wave vector, i.e., $|\mathbf{k}|= 1/2\,(\pi/a)$. The length of the vector from the origin to the surface of the plot is proportional to the value of the eigenfrequency for that propagating direction. In this representation, ideal isotropy would correspond to a sphere. Obviously, the behavior of the designed triclinic chiral metamaterial crystal comes close to this ideal. Finer details can be seen from the false-color plot on the surface. (b) On the left, we depict the relative frequency splitting, $2(f_2-f_1)/(f_2+f_1)$, in the same manner. In the middle, we show the ratio, $\zeta$, of the major and minor axes of the ellipse formed by the displacement vector of the first transverse eigenmode, averaged over one unit cell. We false-color code $\zeta$ onto the surface of a sphere. $\zeta=1$ corresponds to circular polarization and $\zeta=0$ to linear polarization. Elliptical polarization leads to a value of $\zeta$ in between $0$ and $1$. To visualize the eigenmodes, we show their displacement vector trajectory in a plane perpendicular to the wave vector, again averaged over one unit cell, on a sphere. Black and red trajectories are for $i=1$ and $i=2$, respectively. Very nearly circularly polarized phonon modes are observed for all propagating direction in three dimensions.}
\label{Figure6}
\end{center}
\end{figure}

Classical Cauchy continuum theory completely neglects chiral behavior. This means that anisotropy measures based on Cauchy elasticity \cite{ranganathan2008universal, fang2019energy} cannot be used for acoustical activity of chiral crystals. Generalized continuum theory is needed to understand the elastic properties of the designed chiral metamaterial crystal from an effective-medium perspective \cite{eringen1966linear}. Our previous study has shown that micropolar continuum theory is sufficient to characterize the chiral effects in cubic chiral metamaterials \cite{chen2020mapping}. In micropolar continuum theory \cite{eringen1966linear}, the splitting between the two transverse bands can be attributed to two factors, anisotropy and chirality. In the long-wavelength limit, the chirality induced splitting vanishes \cite{chen2020mapping} and only the unwanted anisotropy induced splitting remains. To isolate the wanted isotropic chiral behavior, the relative splitting in the limit of zero wave number should be as small as possible. In Fig.\,5(b), we plot the relative splitting, $2(f_2-f_1)/(f_2+f_1)$, along several chosen directions (cf. Fig.\,5(a)). For all of the chosen directions, we find an extremely small relative splitting in the limit $|\mathbf{k}| \rightarrow 0$. Furthermore, the relative splitting along different wave propagating directions overlaps quit well over a broad range of wave numbers $0 < |\mathbf{k}|/(\pi/a) < 0.6$ and hence a broad range of frequencies. This result again indicates that our designed chiral metamaterial crystal exhibits the desired broadband isotropic acoustical activity behavior. 

The above results have been for specific selected wave propagating directions. Next, we investigate the direction dependence of the eigenfrequencies and eigenmodes to further verify the isotropic behavior of our designed chiral metamaterial crystal. In the first row of Fig.\,6, we plot the first three eigenfrequencies, again for a fixed length of the wave vector $|\mathbf{k}|= 1/2\,(\pi/a)$ as in Fig.\,4. For any point on the surface, its distance to the origin is proportional to the eigenfrequency corresponding to wave vector along that direction. The three surfaces are remarkably similar to ideal spheres, indicating a very nearly isotropic behavior. For instance, the variation of the first eigenfrequency with respect to the mean value versus direction is less than $\pm 0.36\%$, with a minimum of $112.26\,\rm{kHz}$ and a maximum of $113.07\,\rm{kHz}$. For the most interesting relative frequency splitting, $2(f_2-f_1)/(f_2+f_1)$, we again obtain a nearly ideal spherical surface. To more quantitatively characterize the circular polarization of the transverse eigenmodes, we define the dimensionless quantity $\zeta$. $\zeta$ is the ratio of the major and minor axes of the ellipse described by the tip of the displacement vector of the first transverse eigenmode, averaged over one unit cell. Linearly and circularly polarized eigenmodes are represented by $\zeta = 0$ and $\zeta = 1$, respectively. Intermediate values correspond to elliptical polarization. In the middle panel of Fig.\,6(b), we falso-color plot $\zeta$ on the surface of a sphere in momentum space. On the right-hand side of panel (b), we depict the trajectories of the displacement vector for $i=1,\,2$. Apparently, very nearly circularly polarized eigenmodes are observed for all wave vector directions in three dimensions. We emphasize once again that the triclinic chiral metamaterials structure itself does not have four-fold rotational symmetry along the principal cubic directions of the translational lattice. Therefore, our calculations have neither assumed nor implied this symmetry. Nevertheless, the chiral behavior depicted in Fig.\,6 does show three four-fold rotational symmetry axes. Intuitively, this important finding can be interpreted in that the chiral meta-rods can conceptually be replaced by chiral effective-medium rods, the properties of which have uniaxial symmetry. If each meta-rod is conceptually replaced by such an effective-medium uniaxial rod, even the crystal structure recovers simple-cubic symmetry. 

\begin{figure}[htbp]
\begin{center}
\includegraphics[width=15.0cm, angle=0]{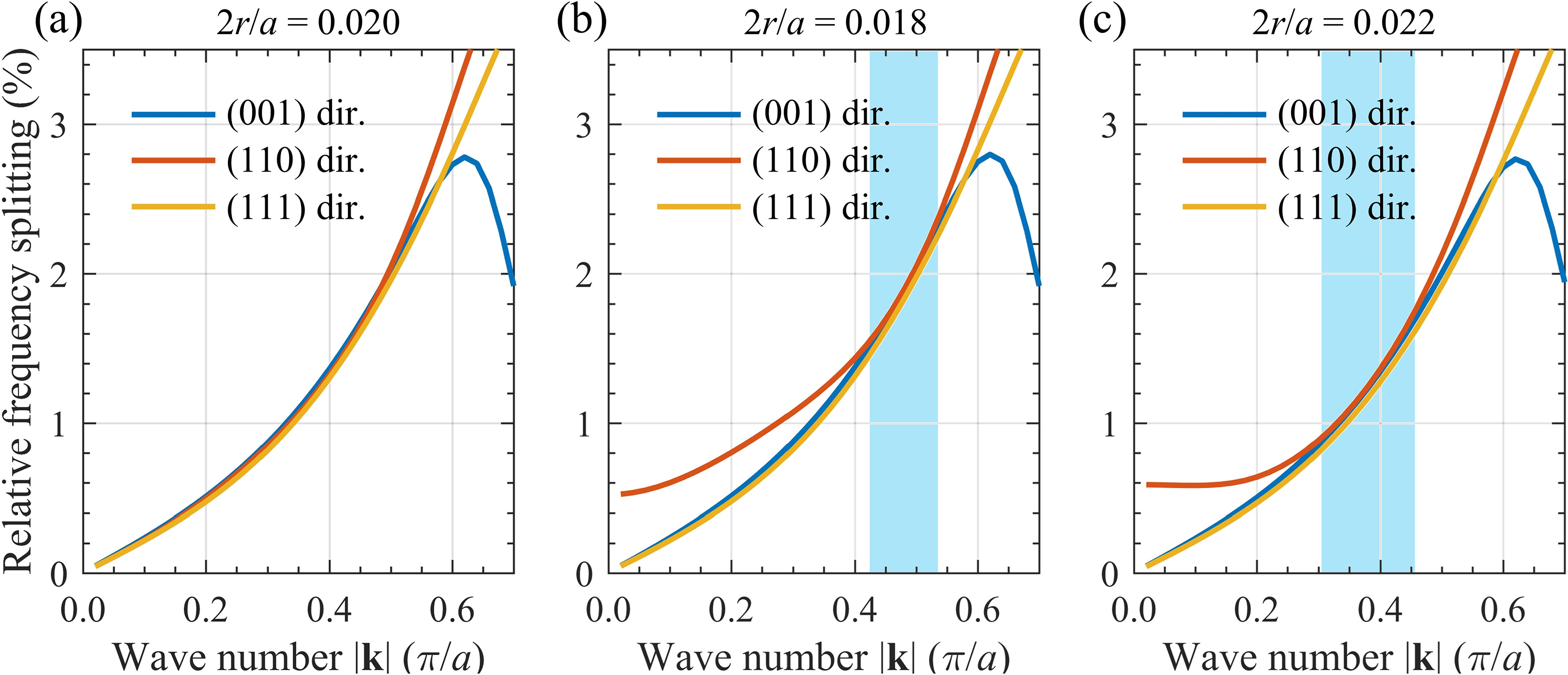}
\caption{Influence of the rod diameter, $2r$, on acoustical activity in the designed chiral metamaterial. The relative frequency splitting, $2(f_2-f_1)/(f_2+f_1)$, along a cubic principal direction (001), a face diagonal direction (110), and a space diagonal direction (111) are shown with the rod diameters: (a) $2r/a=0.020$, (b) $2r/a=0.018$, and (c) $2r/a=0.022$, respectively. For case (a) corresponding to the optimized rod diameter used above, $2r/a=0.02$, broadband isotropic acoustical activity is achieved starting from the long-wavelength limit, whereas accidental degeneracies for the relative frequency splittings are only observed for a small range of wave numbers (shaded in light blue) for the cases (b) and (c). This finding clearly shows that isotropic chiral behavior is obtained only for a certain set of designed geometry parameters of the metamaterial.}
\label{Figure7}
\end{center}
\end{figure}

Finally, we investigate the robustness of the chiral behavior versus the design parameters. As an example, we consider variations of the relative rod diameter $2r/a$. Such variations may, for example, come about from imperfections in a 3D additive manufacturing process. Following our previous analysis, we evaluate whether isotropic acoustical activity remains by the proxy of the relative frequency splitting between the two transverse bands along the (001), (110), and (111) directions. When the rod diameter is decreased (cf. Fig.\,7(b)) or increased (cf. Fig.\,7(c)) by $10\%$ relative to the optimized value of $2r/a=0.02$ (cf. Fig.\,7(a)), the relative frequency splitting for the three directions no longer overlaps from near zero wave numbers upwards. Instead, we find accidental degeneracy only within a specific range of wave numbers (shaded in light blue), which leads to narrow-band isotropic acoustical activity. This means that the chiral behavior at elevated frequencies is fairly robust against variations of $2r/a$, but relative deviations exceeding $\pm 10\%$ should certainly be avoided. The other geometry parameters, such as $R/L$ and $d/L$, have a similar qualitative influence (not depicted).

\section{Conclusions}
By tailoring accidental degeneracies in reciprocal space, we have rationally designed and characterized theoretically 3D triclinic chiral truss-based metamaterial crystals exhibiting isotropic chiral phonons and isotropic acoustical activity. The resulting polarization-rotation power is comparable in magnitude to that of previous highly anisotropic cubic metamaterial crystals. The behavior can be well described by Eringen micropolar continuum elasticity. This novel blueprint bring us much closer to attractive future experimental realizations of such microstructured chiral metamaterial crystals by means of advanced 3D additive manufacturing. In passing, we have also rationally designed achiral 3D simple-cubic truss-based metamaterial crystals with isotropic elastic properties.

\section*{Acknowledgements}
We thank Tobias Frenzel (KIT) and Julian K\"opfler (KIT) for stimulating discussions.

\section*{Funding}
Y. C. acknowledges support by the Alexander von Humboldt Foundation and by the National Natural Science Foundation of China (Contract No. 11802017). This research has additionally been funded by the Deutsche Forschungsgemeinschaft (DFG, German Research Foundation) under Germany’s Excellence Strategy via the Excellence Cluster 3D Matter Made to Order (EXC-2082/1-390761711), which has also been supported by the Carl Zeiss Foundation through the Carl-Zeiss-Foundation-Focus@HEiKA, by the State of Baden-Württemberg, and by the Karlsruhe Institute of Technology (KIT). We further acknowledge support by the Helmholtz program Science and Technology of Nanosystems (STN), and by the associated KIT project Virtual Materials Design (VIRTMAT). M. K. is grateful for support by the EIPHI Graduate School (Contract No. ANR-17-EURE-0002) and by the French Investissements d’Avenir program, project ISITEBFC (Contract No. ANR-15-IDEX-03).

%

\end{document}